\documentclass[reprint,superscriptaddress,aps,prl,]{revtex4-2}
\usepackage{amsmath}
\usepackage{amssymb}
\usepackage{graphicx}
\usepackage{epstopdf}
\usepackage[colorlinks=true]{hyperref}
\usepackage{physics}
\usepackage{mathrsfs}
\usepackage{comment}

\renewcommand\({\begin{equation}}	
\renewcommand\){\end{equation}}
\renewcommand\[{\begin{eqnarray}}	
\renewcommand\]{\end{eqnarray}}
\newcommand{\al}[1]{\begin{aligned}#1\end{aligned}}
\usepackage[dvipsnames]{xcolor}

\begin{document}

\title{Dissipative Spin-wave Diode and Nonreciprocal Magnonic Amplifier}

\author{Ji Zou}
\author{Stefano Bosco}
\author{Even Thingstad}
\author{Jelena Klinovaja}
\author{Daniel Loss}
\affiliation{Department of Physics, University of Basel, Klingelbergstrasse 82, 4056 Basel, Switzerland}

\begin{abstract}
We propose an experimentally feasible \textit{dissipative} spin-wave diode comprising two magnetic layers coupled via a non-magnetic spacer.  We theoretically demonstrate that the spacer mediates not only coherent interactions but also dissipative coupling. Interestingly, an appropriately engineered dissipation  engenders a nonreciprocal device response, facilitating the realization of a spin-wave diode. This diode permits wave propagation in one direction alone, given that the coherent Dzyaloshinskii-Moriya (DM) interaction is balanced with the dissipative coupling.
The polarity of the diode is determined by the  sign of the DM interaction. Furthermore, we show that when the magnetic layers undergo incoherent pumping,  the device operates as a unidirectional spin-wave amplifier. 
The amplifier gain is augmented by cascading multiple magnetic bilayers.  By extending our model to a one-dimensional ring structure, we  establish a connection between the physics of spin-wave amplification and non-Hermitian topology.   Our proposal opens up a new avenue for harnessing inherent dissipation in spintronic applications.
\end{abstract}

\date{\today}
\maketitle

\textit{Introduction.}|
A main theme of magnonics is the utilization of spin waves, or their quanta magnons, for information processing and transmission,  to develop innovative computing and communication technologies~\cite{Chumak:2015aa,YUAN20221,chumak2021roadmap,lenk2011building,barman20212021,nikitov2015magnonics,yu2021magnetic}. 
The ability to directionally control information transmission is a cornerstone in both classical and quantum information processing systems~\cite{tooley2019electronic,crecraft2002analog,rymarz2021hardware}. This control is facilitated through the employment of essential components such as diodes.
In magnonics, a spin-wave diode, which permits the unidirectional transmission of spin waves, holds critical importance. Such a device can function as a spin-wave rectifier and foster the development of more efficient and adaptable spintronic devices, such as memory devices, sensors, and spin-wave-based logic circuits.
It   stimulated numerous recent studies on nonreciprocal couplings and chiral spin waves, both theoretically~\cite{yu2019chiral,yu2022efficient,yuan2023unidirectional,udvardi2009chiral} and experimentally~\cite{gitgeatpong2017nonreciprocal,wang2018unidirectional,yu2023chirality,di2015direct}. Several  proposals for spin-wave diodes have emerged~\cite{van2011rectification,nakata2017spin,lan2015spin,szulc2020spin,grassi2020slow,shichi2015spin}, relying on anisotropic exchange,  dipolar or Dzyaloshinskii-Moriya (DM) interactions. These proposals share one common feature|they build upon  purely coherent interactions in the system, requiring low levels of damping for high diode efficiency.

In this work,  we provide a different paradigm for realizing a spin-wave diode that leverages \textit{dissipative} couplings in the system, which allows us to   achieve a perfect spin-wave diode even when the  damping is comparable to the coherence interactions~\cite{metelmann2015nonreciprocal}.  We consider an experimentally feasible ferromagnetic  bilayer structure separated by a non-magnetic spacer layer as depicted in Fig.~\ref{fig1}(a). We demonstrate that this spacer not only mediates coherent couplings such as the DM interaction between two magnetic layers but also gives rise to a  dissipative coupling and local damping. Interestingly, we find that an ideal diode, facilitating unidirectional transmission while completely blocking the opposite direction, can be achieved  in the presence of significant dissipative coupling, when it is  balanced with the DM interaction. We thus refer to such a system as a \textit{dissipative} spin-wave diode.

\begin{figure}[t!]
	\centering\includegraphics[width=\linewidth]{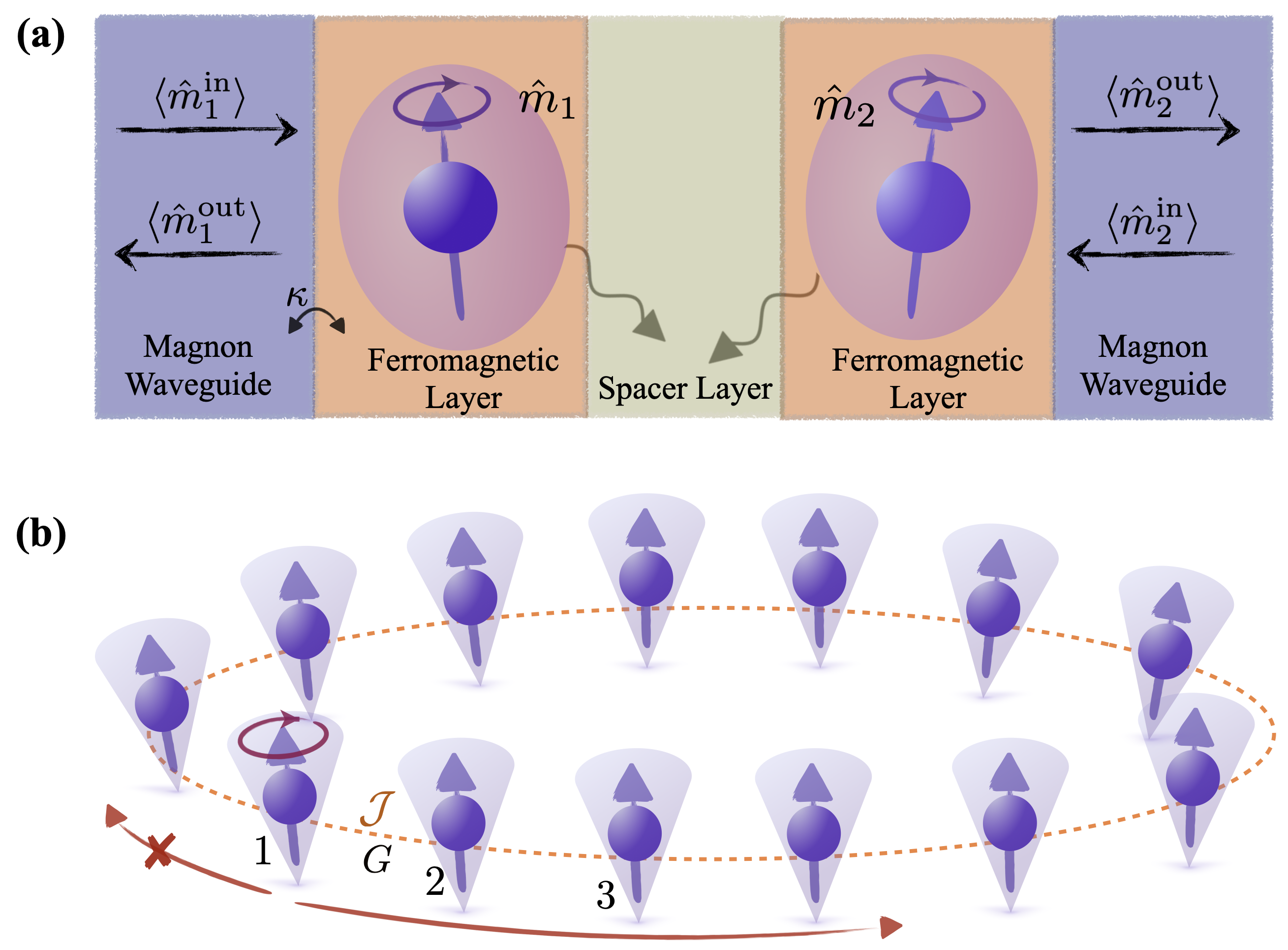}
 \caption{(a) Schematic diagram of the \textit{dissipative} spin-wave diode realized by a magnetic bilayer coupled through a non-magnetic spacer. When the coherent and dissipative couplings mediated by the spacer layer between the two magnetic layers are balanced, the spin wave can only propagate in one direction, while being blocked in the opposite direction, realizing a spin-wave diode. (b) Generalization to many magnetic layers coupled to their neighboring sites via both coherent and dissipative couplings, $\mathcal{J}$ and $G$. When these couplings are balanced, the launched spin wave in the system can only travel unidirectionally.}
  \label{fig1}
\end{figure}

We further demonstrate that when the magnetic layers are incoherently pumped, nonreciprocal amplification of spin waves can occur. In this scenario, the unidirectional spin wave signal is  enhanced. This type of amplifier has the potential for broad applications in magnonics. It  is capable of effectively filtering out thermal noise and  also  enables the detection of weak signals while simultaneously protecting them from unwanted backaction caused by the read-out devices~\cite{reichenbach2011unidirectional,wen2022unidirectional,doornenbal2019spin,petite1980observation}.
The gain of the amplifier can be  enhanced through the cascading arrangement of several magnetic bilayers.
We thus extend our study to an array of magnetic layers, where neighboring sites are coupled through both coherent and dissipative couplings.  When these couplings are balanced, the spin waves launched in the system can only propagate unidirectionally, as depicted in Fig.~\ref{fig1}(b). We further establish a connection between  amplification  and the non-Hermitian topology of the dissipative magnetic system~\cite{yu2023non,hurst2022non,epyaroslav,flebus2020non,deng2022non}. Our findings suggest a new and promising approach for utilizing the inherent dissipation in magnetic systems for spintronic applications.

\textit{Model and mean-field dynamics.}|We model the magnetic bilayer system  sketched in Fig.~\ref{fig1}(a) with the following Hamiltonian: 
\(  \hat{H}=\hat{H}_M+\hat{H}_E+\hat{H}_{\text{ME}},    \)
where $\hat{H}_M=-b(\hat{S}_1^z + \hat{S}_2^z)$ governs  the dynamics of the two magnetic layers described by macroscopic spins $\hat{S}_i$ in an applied magnetic field $b\hat{z}$. We leave the Hamiltonian of the spacer layer $\hat{H}_{E}$ unspecified. 
 The coupling between the magnetic and spacer layers is captured by $\hat{H}_{\text{ME}}= \lambda \sum_{i=1,2} ( \hat{S}_i^- \hat{E}_i^- +\text{h.c.} )$ with $\hat{S}^\pm \equiv \hat{S}^x \pm i \hat{S}^y$ and coupling strength $\lambda$. Here $\hat{E}_i^\pm$ are operators for quasiparticles, such as phonons,  within the spacer, that act as decay channels for the magnons in magnetic layers. After performing the Holstein-Primakoff transformation~\cite{PhysRev.58.1098} and keeping leading terms, we arrive at $\hat{H}_M\approx\hbar \Omega \sum_{i=1,2} \hat{m}^\dagger_i \hat{m}_i$ and $\hat{H}_{\text{ME}}=\sum_{i=1,2}  ( \hat{m}_i^\dagger \hat{E}_i^- +\text{h.c.} )$.  Here, $\hat{m}_i$, $i=1, 2$, are magnon operators for two magnetic layers, obeying the standard bosonic algebra $[\hat{m}_i, \hat{m}_j^\dagger]=\delta_{ij}$.  
The ferromagnetic resonance frequency, denoted as $\Omega$ and typically residing within the GHz regime (1-100 GHz), is proportional to the applied magnetic field, and in general, it can also be set by an easy-axis anisotropy.  We have  absorbed constant prefactors into the definition of $\hat{E}_i^\pm$ for notational convenience.

To examine the dynamics of the two magnetic layers without monitoring the dynamics of the spacer, we trace out the spacer degree of freedom~\cite{Heinz} and obtain the following master equation for the (reduced) density matrix $\hat{\rho}$ for two magnon modes at zero temperature~\cite{sm_diode}: 
\( \dv{}{t} {\hat{\rho}}=-\frac{i}{\hbar} [\hat{H}_M+ \hat{H}_{C} , \hat{\rho}] +\sum_{k, j =1,2} \mathcal{L}_{kj}^\downarrow \hat{\rho}, \label{eq:2} \)
where the spacer-mediated coherent interaction reads: 
\( \hat{H}_C\!=\! \mathcal{J} \hat{m}_1^\dagger \hat{m}_2 + \mathcal{J}^* \hat{m}_2^\dagger \hat{m}_1 , \;\;\text{with}\;\; \mathcal{J}\! =\! \frac{ \mathcal{G}^R_{12}(\Omega) +\mathcal{G}^A_{12}(\Omega)  }{2\hbar} .\)
Here, the retarded and advanced Green's functions follow their standard definitions~\cite{bruus} $G^{R, A}_{12} (t) = \mp i \Theta(\pm t) \langle [\hat{E}_1^-(t), \hat{E}_2^+]\rangle$~\cite{diode_average,sm_diode}  and we use: $\mathcal{G}(\omega)=\int d\tau \, e^{i\omega \tau} G(\tau)$.  The coherent coupling $\mathcal{J}$ is complex-valued in general. Its real part stands for the symmetric exchange between the two magnetic layers, while the imaginary component represents the   DM interaction, which is non-vanishing only when the spatial inversion symmetry of the spacer is broken~\cite{sm_diode}. We note that both  are highly tunable in experiments (capable of reaching MHz regime), for instance by adjusting the thickness of the spacer layer~\cite{parkin_1990_prl,parkin_1991_prl,husain2022large,yun2023anisotropic} or applying an electric field~\cite{srivastava2018large,fillion2022gate,koyama2018electric,vedmedenko2019interlayer}. This tunability should enable the experimental realization of a dissipative spin-wave diode.

The second term in the master equation~\eqref{eq:2} accounts for the non-unitary (dissipative) evolution of the magnetization due to the spacer layer, with the following Lindblad form: $ \mathcal{L}_{kj}^\downarrow \hat{\rho}  =\gamma_{kj}^\downarrow \big[  \hat{m}_j  \hat{\rho} \hat{m}^\dagger_k - \frac{1}{2}\{ \hat{m}_k^\dagger \hat{m}_j, \hat{\rho} \} \big]$. The dissipative parameters are given by $\gamma_{kj}^\downarrow =i\mathcal{G}^>_{kj}(\Omega)/\hbar^2, $
where ${G}_{kj}^>(t)\equiv -i \langle \hat{E}_k^-(t) \hat{E}_j^+ \rangle$ is the greater Green's function~\cite{bruus}. Here $\gamma\equiv \gamma_{jj}^\downarrow$ is real-valued by its definition~\cite{sm_diode} and represents the local magnon decay, leading to the Gilbert damping. It falls within the MHz regime when the induced Gilbert damping is approximately $10^{-3}$.
Further, $G\equiv \gamma_{21}^\downarrow$ represents non-local damping of magnons in the two magnetic layers, and we refer to it as the \textit{dissipative} coupling~\cite{zou2022prb}. 
This non-local  damping can be comparable in magnitude to the local one in experiments~\cite{heinrich2003dynamic,subedi2023magnon}.
 It is in principle also tunable, as it is governed by the response function in the spacer, similar to the coherent coupling. 
 Such dissipative coupling is a key ingredient in recent proposals for non-Hermitian topological magnonic phases~\cite{hurst2022non,epyaroslav,flebus2020non,deng2022non}, and has also been investigated in cavity magnonic~\cite{harder2018level,wang2019nonreciprocity,xu2019cavity}   and photonic~\cite{metelmann2015nonreciprocal,wang2023quantum,metelmann2017nonreciprocal} systems.
Its real and imaginary components correspond to the dissipative symmetric and DM-like interactions, respectively. The latter emerges only when the inversion symmetry is broken~\cite{sm_diode}. In the absence of spin pumping,  the system is in its natural thermodynamic equilibrium and the dissipative coupling is constrained by the local damping $|G|\leq \gamma$~\cite{zou2022prb}, which ensures the complete positivity of the system dynamics. While we have assumed  zero temperature for simplicity,  a general discussion of finite temperature is provided in the SM~\cite{sm_diode}, which does not alter our conclusion as long as it is small compared to the resonance  frequency $\Omega$. 

We obtain the mean-field dynamics of two magnetic layers from the master equation~\eqref{eq:2} by evaluating:
\( \dv{}{t} \langle \hat{m}_i \rangle = -\frac{i}{\hbar}\langle [\hat{m}_i, \hat{H}_M+\hat{H}_C]\rangle +\tr\big[ \hat{m}_i \sum_{k,j} \mathcal{L}_{kj}^\downarrow \hat{\rho} \big], \)
which yields  $i\,\text{d}\vb* \psi /\text{d}t=\mathcal{H}\vb* \psi $ with $\vb* \psi\equiv (\langle \hat{m}_1\rangle, \langle \hat{m}_2\rangle  )^T$ and an \textit{effective}  {non-Hermitian} Hamiltonian~\cite{sm_diode}:
\(   \mathcal{H}=\mqty[ \Omega- i \gamma/2 &  \mathcal{J}/\hbar-iG^*/2   \\  \mathcal{J}^*/\hbar - i G/2   & \Omega-i\gamma/2  ].   \label{eq:5} \)
The non-Hermitian character reflects  the dissipative nature of the magnetic dynamics.
Here,   $\langle \hat{m}_i \rangle\equiv \tr [\hat{m}_i \,\hat{\rho}  ]\propto \langle\hat{S}^x_i\rangle + i \langle \hat{S}^y_i\rangle$ is the in-plane spin component. We note that the phase of the dissipative coupling $G$ can be gauged away and absorbed into the definition of the coherent coupling $\mathcal{J}$ [only the relative phase $\Phi\equiv \text{arg}(\mathcal{J}/G^*)$ matters]. For this reason, we assume that $G$ is  positive-valued  and  $\mathcal{J} = e^{i\Phi} |\mathcal{J}|$, henceforth. We stress that a non-zero value of $\Phi$ indicates the presence of a finite effective DM interaction that breaks spatial inversion symmetry.

\textit{Dissipative spin-wave diode.}|To investigate the unidirectional transmission of spin waves, we couple magnon waveguides to the bilayer system, as depicted in Fig.~\ref{fig1}(a), with a coupling rate of $\kappa$. The response of the system is described by the Green's function~\cite{bruus}: 
\( \hat{G}(\omega ) = \frac{1}{\omega-\mathcal{H}},  \label{eq:6} \)
where we replace the local damping $\gamma$ with $\Gamma = \gamma + \kappa$ in the Hamiltonian~\eqref{eq:5} to account for the additional channel for magnon leakage provided by the magnon waveguide. 
We note that the magnitudes of the off-diagonal elements of $\mathcal{H}$ are different  if  firstly  the effective DM interaction is finite ($\Phi\neq 0, \pi$), and secondly  the dissipative coupling is non-vanishing ($G\neq 0$). This signals the emergence of a nonreciprocal phenomenon in the bilayer structure. Of particular interest is the scenario where $|\mathcal{H}_{12}|=0$ but $|\mathcal{H}_{21}|> 0$. This is  achieved when the following condition is fulfilled:
\(   \mathcal{J}/\hbar =iG/2 \label{eq:7}\, .\)
Here, the dissipative coupling $G$ is balanced with the  coherent coupling $|\mathcal{J}|$, which  is purely DM interaction ($\Phi=\pi/2$).   The scattering matrix, related to the Green's function via $\hat{S}(\omega)=1-i\kappa \hat{G}(\omega)$ as dictated by the input-output theory~\cite{gardiner1985input,clerk2010introduction,clerk2022introduction}, at resonance frequency reads: 
\( \hat{S}(\Omega)= \mqty[   \dfrac{\gamma-\kappa}{\gamma+\kappa}  & 0 \\ &   \\ \dfrac{4\kappa G}{(\gamma+\kappa)^2}  &    \dfrac{\gamma-\kappa}{\gamma+\kappa}     ]. \label{eq:8}  \)
It relates the incoming and outgoing spin waves as shown in Fig.~\ref{fig1}(a), $\vb* \psi^{\text{out}}=\hat{S}(\omega) \vb* \psi^{\text{in}}$, with $\vb*\psi^\text{in} \equiv ( \langle \hat{m}_1^{\text{in}} \rangle,  \langle \hat{m}_2^{\text{in}} \rangle )^T.$ We point out  that the non-unitarity observed in this scattering matrix arises from dissipation, while its asymmetry leads to the nonreciprocal behavior.

\begin{figure}
	\centering\includegraphics[width=\linewidth]{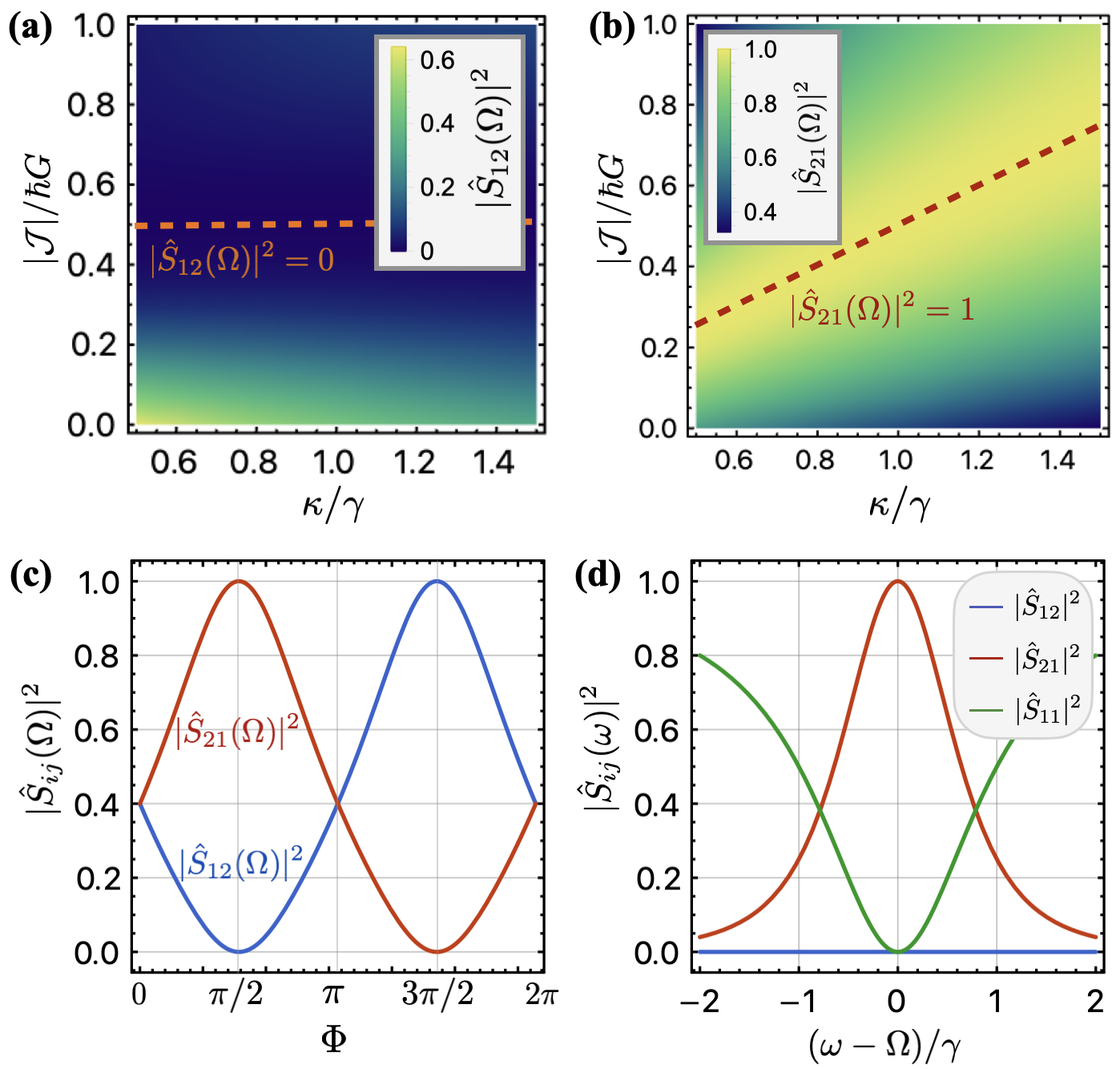}
 \caption{Performance of a dissipative spin-wave diode. 
 (a) The transmission coefficient $|\hat{S}_{12}(\Omega)|^2$ is plotted as a function of the coupling ratio $|\mathcal{J}|/\hbar G$ and the damping ratio $\kappa/\gamma$ with $\Phi=\pi/2$.
An ideal diode is attained when   the coherent and dissipative couplings are balanced  (depicted by the dashed line), $|\mathcal{J}|=\hbar G/2$,  independent of the ratio $\kappa/\gamma$.
  (b) The transmission coefficient $|\hat{S}_{21}(\Omega)|^2$ is shown as a function of $|\mathcal{J}|/\hbar G$ and  $\kappa/\gamma$ with $\Phi=\pi/2$. The dashed red line indicates the condition for perfect transmission, $|\hat{S}_{21}(\Omega)|^2=1$, which is achieved when $\kappa=2|\mathcal{J}|/\hbar$. 
 (c) Transmission coefficients as a function of $\Phi$.  At $\Phi=\pi/2$, $3\pi/2$, where the coherent coupling is purely DM interaction, we observe a perfect dissipative diode with its direction dictated by the sign of the DM interaction. 
 (d) Transmission and reflection coefficients as a function of spin-wave frequency with $\Phi=\pi/2$. The system acts as a diode for all frequencies and achieves a perfect diode effect at resonance $\omega=\Omega$. In plots (c) and (d), we set $|\mathcal{J}|/\hbar=G/2$ and $\kappa=\gamma$.   In all plots, we use $\gamma=G$~\cite{sm_diode}.}
  \label{fig2}
\end{figure}

We observe that left-propagating spin waves are entirely blocked, with a vanishing transmission coefficient $|\hat{S}_{12}|^2=0$. In contrast, waves traveling in the opposite  maintain a finite transmission coefficient,  realizing a perfect diode.
We point out that there is no amplification as $|\hat{S}_{21}|^2\leq 1$  without spin pumping. It should also be noted that reflected spin waves are generally observed, with $\hat{S}_{11}\neq 0$. However, when the two magnon-decay channels (the magnon waveguide and the spacer)  are matched $\gamma=\kappa$, the reflection coefficient vanishes, and the transmission coefficient reaches its maximal value $|\hat{S}_{21}|^2=(G/\gamma)^2$ (a perfect transmission of spin wave is achieved when $G=\gamma$). 

While  condition \eqref{eq:7} enables an ideal dissipative spin-wave diode, it is worth noting that a well-performing diode, with $|\hat{S}_{21}|^2\approx 1$ and $|\hat{S}_{12}|^2\ll 1$, can be attained without requiring fine-tuning of the parameters. 
We depict the  transmission coefficients for the two opposite directions as functions of the ratio of the two couplings $|\mathcal{J}|/\hbar G$ and the ratio of the two dampings $\kappa/\gamma$ in Fig.~\ref{fig2}(a) and (b). 
We observe that a large transmission coefficient in one direction  and a small one in the opposite direction can be obtained over a wide range of parameter values. Besides, a perfect unidirectional spin-wave blockade, transmitting sizable signals in the opposite direction, can be accomplished over a bandwidth set by local damping $\gamma$  (MHz regime), as depicted in Fig.~\ref{fig2}(d).

We  emphasize that the directionality of the spin-wave diode is determined by the sign of the effective DM interaction $\mathcal{D}=|\mathcal{J}|\sin\Phi$. Specifically, when $\Phi=\pi/2$ [necessary to satisfy condition~\eqref{eq:7}], spin waves propagating from right to left are blocked, while  opposite traveling  waves are blocked when the sign of $\mathcal{D}$ is flipped ($\Phi=3\pi/2$).
This is illustrated in Fig.~\ref{fig2}(c). For $\Phi=0$, $\pi$, when the DM interaction vanishes and the inversion symmetry is preserved, the device is reciprocal and $|\hat{S}_{12}|=|\hat{S}_{21}|$.

\textit{Unidirectional amplification of spin wave.} Applying spin-transfer~\cite{slonczewski1996current,berger1996emission} or spin Seebeck~\cite{bauer2012spin} torques allows for the local pumping of magnetic layers out of their natural thermodynamic equilibrium. We leverage this to explore a broader range of experimentally tunable parameters, examining the possibility of novel dynamical phenomena.
 The spin pumping can be modeled as an additional term in the master equation~\eqref{eq:2}~\cite{chelpanova2021intertwining}: $\mathcal{L}^{\text{pump}}\hat{\rho}=\gamma^\uparrow\sum_{i=1,2}\big[ \hat{m}_i^\dagger \hat{\rho}\hat{m}_i - \frac{1}{2} \{ \hat{m}_i \hat{m}_i^\dagger, \hat{\rho}   \}  \big] $ with  pumping rate $\gamma^\uparrow>0$. This pumping leads to a reduction of the local  damping in the effective Hamiltonian~\eqref{eq:5}~\cite{sm_diode}, where we now replace $\gamma$ with $\tilde{\gamma}\equiv \gamma-\gamma^\uparrow$. We point out that our system overall is still damped with  effective damping parameter $\tilde{\gamma}>0$. 
 
To see how the pumping modifies the behaviour of the spin wave diode, we consider the scenario where the condition~\eqref{eq:7} is satisfied and the local damping is matched $\kappa=\tilde{\gamma}$ for simplicity. The scattering matrix for spin waves at resonance then reads: 
\( \hat{S}(\Omega)=\dfrac{G}{\gamma-\gamma^{\uparrow}} \mqty[  \; 0 & 0\;   \\ \; 1 & 0 \; ].   \)
We observe that spin pumping allows the realization of an unidirectional amplifier, where the spin wave propagation is hindered in one direction, and the  signal is enhanced   in the opposite direction, when $G >\gamma-\gamma^\uparrow$. This feature can find broad applications in {the next-generation of} spin-wave-based devices~\cite{reichenbach2011unidirectional,wen2022unidirectional,doornenbal2019spin,petite1980observation}. We remark that the linear treatment breaks down when $G \gg \gamma-\gamma^\uparrow$, and in the case of zero effective damping ($\gamma=\gamma^\uparrow$), the system exhibits instability.

We  emphasize that the unidirectional amplification is effective over a wide range of frequencies, as shown in Fig.~\ref{fig3}. The dashed orange curve corresponds to the boundary that separates the parameter regimes with and without amplification, which is determined by $|\omega-\Omega|<\sqrt{ \tilde{\gamma} (G-  \tilde{\gamma})}$.  Observably, the amplification bandwidth initially expands as the system undergoes pumping. However, once the effective damping diminishes to $\tilde{\gamma}=G/2$, the bandwidth begins to contract (at this point, the maximal bandwidth is  $\sim G$, within the MHz regime). 

 \begin{figure}[t]
	\centering\includegraphics[width=0.68\linewidth]{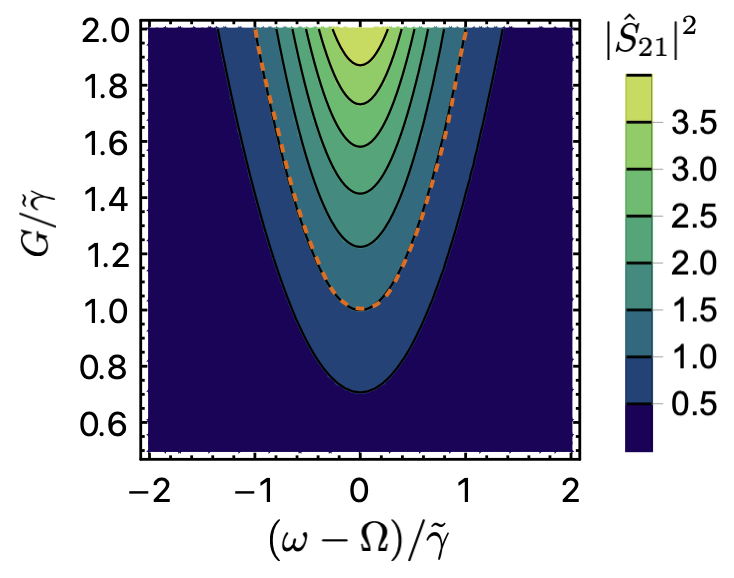}
 \caption{Unidirectional amplification in the presence of spin pumping. The transmission coefficient $|\hat{S}_{21}|^2$ is shown as a function of frequency $\omega$ and the ratio of dissipative coupling $G$ to effective local damping $\tilde{\gamma}$. Without spin pumping $G \leq \tilde{\gamma}$, no amplification is observed. With spin pumping, the effective local damping $\tilde{\gamma}$ can be reduced, allowing  $G>\tilde{\gamma}$ and resulting in amplification. The dashed orange curve encloses the region with unidirectional amplification.  We have assumed  $\kappa=\tilde{\gamma}$ and condition~\eqref{eq:7} to get $\hat{S}_{12}=0$.   }
  \label{fig3}
\end{figure}

\textit{Cascading multiple magnetic  layers.}|Employing a cascade of multiple magnetic bilayers can  enhance the gain of the amplifier. Here we consider a ring structure for concreteness, as sketched in Fig.~\ref{fig1}(b), where adjacent sites  coupled to each other via both coherent coupling $\mathcal{J}$ and dissipative coupling $G$ mediated by spacer layers between them. We describe the system with the following effective non-Hermitian Hamiltonian~\cite{diode_hamiltonian}:  
\(
 \al{\mathcal{H} \!=\!  &\sum_j \hbar (\Omega - i\gamma) \hat{m}^\dagger_j \hat{m}_j \! +\!\sum_j (\mathcal{J} -i\hbar G/2 ) \hat{m}_j^\dagger  \hat{m}_{j+1} 
         \\ &  + \sum_j (\mathcal{J}^* - i\hbar G/2) \hat{m}^\dagger_{j+1} \hat{m}_j ,}   \label{eq:10}  \)
which in momentum space takes the diagonal form $\mathcal{H}= \sum_k h(k) \hat{m}_k^\dagger \hat{m}_k$ where $h(k)=\hbar (\Omega - i\gamma)  + (\mathcal{J} -i\hbar G/2 ) e^{ik} + (\mathcal{J}^* - i\hbar G/2)  e^{-ik}$. The central quantity that governs the response of the system is the Green's function, which can be expressed as~\cite{sm_diode}
\( \bra{l}\hat{G}(\omega)\ket{j}= \oint_{|z|=1} \frac{dz}{2\pi i} \frac{z^{l-j}}{ z [\omega -h(z)/\hbar ]},    \label{eq:11}  \)
in real and frequency space. Here, $z=e^{ik}$~\cite{diode_z} and the integral can be  evaluated by using Cauchy's residue theorem~\cite{xue2021simple}. As expected, Eq.~\eqref{eq:6} is recovered when we calculate the Green's function for two adjacent sites with Eq.~\eqref{eq:11}~\cite{sm_diode}.

Considering again condition~\eqref{eq:7} to be satisfied, the second term in Hamiltonian~\eqref{eq:10} vanishes, indicating that magnons cannot travel clockwise. This is also reflected in the Green's function $\bra{j}\hat{G}(\omega) \ket{j+l} =0 $, which vanishes for  $l>0$.  Hence, a spin wave launched in the system can only propagate in the counterclockwise direction, as shown in Fig.~\ref{fig1}(b).  Evaluating the corresponding Green’s function, we find that  $\bra{j+l}\hat{G}(\omega)\ket{j} \sim \alpha^l$  for $l>0$, where $\alpha=G/(\omega-\Omega-i\gamma)$~\cite{sm_diode}.
 In the absence of spin pumping $G\leq \gamma$, the unidirectional spin wave in the system cannot be amplified for all frequencies since $|\alpha|\leq 1$. 

On the other hand, when the system is pumped and $|\alpha|>1$, the magnetic ring in Fig.~\ref{fig1}(b) experiences an instability, where the spin wave keeps accumulating energy while circulating in the system, resulting in an unbounded growth of signal and a breakdown of the linearized treatment.
The presence or absence of such amplification can be characterized through a topological index associated with the complex spectrum $h(k)$ of the system~\cite{wanjura2020topological}.  Introducing a planar vector field $\vb n (k)=(\Im[\omega-h(k)], \Re[\omega-h(k)])/\hbar$~\cite{sm_diode}, we define a winding number $\mathcal{N}\equiv (1/2\pi) \int_0^{2\pi}dk\, \hat{z} \cdot (\vb n \times \partial_k \vb n)$.
This winding number takes the value $\mathcal{N}=0$ when  $|\alpha|<1$ and the spin wave decays, and $\mathcal{N}=1$ when $|\alpha|>1$ and the signal grows exponentially. The amplification of spin waves thus may serve as an experimental indicator of a topologically non-trivial magnonic phase~\cite{brunelli2022restoration}. 

\textit{Conclusion.}|We investigated a magnetic bilayer system and utilized a master equation approach to show that the spacer layer can mediate both coherent and dissipative couplings. When the resulting  DM interaction is balanced with the dissipative coupling,  we obtain a spin-wave diode. Furthermore, by pumping the two magnetic layers, the spin wave diode can be promoted to a unidirectional spin-wave amplifier.  The gain can be improved by cascading several amplifiers.  We also generalized our analysis to a one-dimensional ring structure,  and identify  a connection between the physics of spin-wave amplification and non-Hermitian topology.

\begin{acknowledgements}
This work was supported by the Georg H. Endress Foundation and by the Swiss National Science Foundation, NCCR SPIN (grant number 51NF40-180604). 
\end{acknowledgements}


\begin{thebibliography}{66}
\bibitem [{dio({\natexlab{a}})}]{diode_average}%
  \BibitemOpen
  \href@noop {} {}  \bibinfo {note} {Here, $\langle
  \hat{A}\rangle$ stands for $ \text{tr}[\hat{\rho}_E\hat{A}]$ where
  $\hat{\rho}_E$ is the thermal state of the spacer.}\BibitemShut {Stop}%
\bibitem [{dio({\natexlab{b}})}]{diode_hamiltonian}%
  \BibitemOpen
  \href@noop {} {} \bibinfo {note} {We can write down the
  Hamiltonian for this ring structure according to Hamiltonian (5). Note that
  we assume $G$ to be real and replace $\gamma$ with $2\gamma$ since each
  magnetic layer is adjacent to two spacer layers.}\BibitemShut {Stop}%
\bibitem [{dio({\natexlab{c}})}]{diode_z}%
  \BibitemOpen
  \href@noop {} {} \bibinfo {note} {The function $h(z)$ is obtained by replacing $e^{\pm ik}$ with $z^{\pm 1}$ in $h(k)$.}\BibitemShut {Stop}%
\end{thebibliography}

%

\onecolumngrid
\clearpage
\setcounter{equation}{0}
\renewcommand{\theequation}{S\arabic{equation}}
\renewcommand{\thefigure}{S\arabic{figure}}
\appendix

{\centering
    \large{\textbf{{Supplemental Material for\\ ``Dissipative Spin-wave Diode and Nonreciprocal Magnonic Amplifier"}}}
\par}

\bigskip

\author{Ji Zou}
\author{Stefano Bosco}
\author{Even Thingstad}
\author{Jelena Klinovaja}
\author{Daniel Loss}
\affiliation{Department of Physics, University of Basel, Klingelbergstrasse 82, 4056 Basel, Switzerland}

\maketitle

In this Supplemental Material, we present (i) master equation for the two magnetic layers at finite temperature, (ii) non-Hermtian Hamiltonian for the mean-field dynamics, (iii) scattering matrix for the spin-wave diode, and (iv) topological index and unidirectional amplification.

\subsection{(i)  Master equation for the two magnetic layers at finite temperature}
In this section, we present the derivation of the master equation for two magnetic layers separated by a non-magnetic spacer layer. The Hamiltonian of the combined system is given by
\( \hat{H}= \hat{H}_M+\hat{H}_E+\hat{H}_{\text{ME}},  \)
where $\hat{H}_M=-b_1 \hat{S}_1^z-b_2 \hat{S}_2^z$ is the bare Hamiltonian  for the two magnetizations of the  magnetic layers with the applied magnetic field $b_i$, which sets the spin quantization $z$-axis. The coupling between the magnetic layers and the spacer layer is described by the Hamiltonian term $\hat{H}_{\text{ME}}\propto \sum_{i=1,2}( \hat{S}_i^- \hat{E}^-_i +\text{h.c.})$. This coupling leads to damping of the spin dynamics through the leakage of magnons into the spacer layer.  We leave the Hamiltonian of the spacer layer $H_E$ unspecified. Since the spin is large, we perform the Holstein–Primakoff transformation and keep the linear terms: $\hat{S}^+\approx \hbar \sqrt{2S}\hat{m}$, $\hat{S}^-\approx \hbar \sqrt{2S}\hat{m}^\dagger$, and $\hat{S}^z=\hbar (S-\hat{m}^\dagger \hat{m})$. With this approximation, we have the following Hamiltonian:
\(  \hat{H}_M\approx \hbar \Omega \sum_{i=1,2} \hat{m}_i^\dagger \hat{m}_i, \;\;\; \text{and}\;\;\; \hat{H}_{\text{ME}}\approx \sum_{i=1,2}(\hat{m}_i^\dagger \hat{E}_i^- + \text{h.c.}),  \label{eq:s2} \)
with $\Omega$ being the  ferromagnetic resonance frequency (assumed to be the same for two layers for simplicity) and we have absorbed prefactors in $\hat{H}_{\text{ME}}$ into the definition of $\hat{E}^\pm$ for notational convenience. It is convenient to work in the interaction picture, where the density matrix of the combined system $\hat{\rho}_{\text{tot}}$ obeys the equation of motion: 
\(  \dv{}{t}\hat{\rho}_{\text{tot}} (t)=\mathcal{L}(t) \hat{\rho}_{\text{tot}} (t).  \)
 The Liouville superoperator is defined as $\mathcal{L}(t) \hat{\rho}_{\text{tot}} (t)\equiv -(i/\hbar) [\hat{H}_{\text{ME}}(t), \hat{\rho}_{\text{tot}} (t)]$ with \(\hat{H}_{\text{ME}}(t)\equiv \text{exp}[i(\hat{H}_M+\hat{H}_E)t] \, \hat{H}_{\text{ME}} \, \text{exp}[-i(\hat{H}_M+\hat{H}_E)t].\)
 To derive the master equation for the (reduced) density matrix $\hat{\rho}$ of the two magnetic layers, we trace out the degrees of freedom within the spacer layer and perform the standard Born and Markov approximations, which yields: 
\(  \dv{}{t}\hat{\rho}(t)=\int^\infty_0 ds \; \tr_E\big [ \mathcal{L}(t) \mathcal{L}(t-s) \hat{\rho}(t)\otimes \hat{\rho}_E   \big],   \label{s5} \)
where we assume that the spacer is in its thermal equilibrium state  $\hat{\rho}_E$.
With the interaction given in Eq.~\eqref{eq:s2}, we can derive the the master equation for the reduced density matrix of the magnons in the following form (in the Schr\"{o}dinger picture): 
\( \dv{}{t} {\hat{\rho}}(t)=-\frac{i}{\hbar} \Big[\hat{H}_M+\mathcal{J}\hat{m}_1^\dagger \hat{m}_2 +\mathcal{J}^*\hat{m}_2^\dagger \hat{m}_1  , \hat{\rho}(t) \Big] +\sum_{i,j=\{1,2\}} \mathcal{L}_{ij}^\downarrow \hat{\rho} (t)+  \sum_{i,j=\{1,2\}} \mathcal{L}_{ij}^\uparrow \hat{\rho} (t),   \)
with the standard Lindbladian defined as: 
\( \sum_{i,j=\{1,2\}} \mathcal{L}_{ij}^\downarrow \hat{\rho}  =\sum_{i,j=\{1,2\}}\gamma^\downarrow_{ij} \big[  \hat{m}_j  \hat{\rho} \hat{m}^\dagger_i - \frac{1}{2}\{ \hat{m}_i^\dagger \hat{m}_j, \hat{\rho} \} \big], \;\;\; \text{and}\;\;\;   \sum_{i,j=\{1,2\}} \mathcal{L}_{ij}^\uparrow \hat{\rho}=  \sum_{i,j=\{1,2\}} \gamma^\uparrow_{ij} \big[ \hat{m}_j^\dagger \hat{\rho} \hat{m}_i - \frac{1}{2}\{ \hat{m}_i \hat{m}_j^\dagger, \hat{\rho}   \}  \big].  \)
Here, we have assume the spacer is U(1) invariant such that correlators $\langle \hat{E}^\pm_i\hat{E}^\pm_j\rangle_T$ vanish. 
We point out that  all coefficients can be expressed in terms of the Green's functions of the spacer layer.
We first focus on the coherent interactions. The coherent coupling $\mathcal{J}$ is given by
\(  \mathcal{J}= \frac{  \mathcal{G}^R_{12}(\Omega) +\mathcal{G}^A_{12}(\Omega)  }{2\hbar },   \)
which is the Eq.~({\color{red}3}) in the main text.  Here, we have used the standard definition of the retarded and advanced Green's functions: 
\( {G}^R_{A,B}(t)\equiv -i\Theta(t)\langle [\hat{A}(t),\hat{ B}] \rangle_T,\;\;\;\text{and}\;\;\; {G}^A_{A,B}(t)\equiv i\Theta(-t)\langle [\hat{A}(t),\hat{ B}] \rangle_T,   \)
where $\Theta(t)$ is the step function. We use the following Fourier transformation: $\mathcal{G}(\omega)=\int d\tau \, e^{i\omega \tau} {G}(\tau)$. Therefore, here we have $\mathcal{G}^R_{12}(\Omega)\equiv -i\int^\infty_0 d\tau e^{i\Omega \tau} \langle [\hat{E}_1^-(\tau), \hat{E}_2^+]  \rangle_T$ with  $\langle \hat{A}\rangle_T \equiv \tr_E(\hat{\rho}_E \hat{A})$. 
We stress that this coupling is complex in general, whose imaginary part represents the Dzyaloshinskii-Moriya (DM)  interaction. It is given by
\( \Im \mathcal{J} \equiv \frac{\mathcal{J}-\mathcal{J}^*}{2i}= \frac{1}{4i\hbar} \big[  \mathcal{G}^R_{12}(\Omega)- \mathcal{G}^R_{21}(\Omega)  +\mathcal{G}^A_{12}(\Omega)- \mathcal{G}^A_{21}(\Omega)   \big],  \)
where we have used the fact that $[\mathcal{G}^R_{AB}(\omega)]^*=\mathcal{G}^A_{B^\dagger A^\dagger}(\omega)$.
It is evident that this DM interaction is non-zero only when the inversion symmetry is broken.

In the dissipative parts, the decay and absorption rates can also be expressed in terms of the Green's functions of the spacer layer:
\(  \gamma^\downarrow_{ij} = \frac{i}{\hbar^2} \mathcal{G}^>_{ij}(\Omega), \;\;\; \text{and}\;\;\; \gamma^\uparrow_{ij}=\frac{i}{\hbar^2} \mathcal{G}^<_{ji}(\Omega),     \)
where $\mathcal{G}^>_{ij}(\Omega) \equiv -i\int d\tau e^{i\Omega \tau} \langle \hat{E}_i^-(\tau) \hat{E}_j^+\rangle_T$ and  $\mathcal{G}^<_{ji}(\Omega) \equiv -i\int d\tau e^{i\Omega \tau} \langle \hat{E}_i^+ \hat{E}_j^-(\tau)\rangle_T$.
The local decay and absorption rates $\gamma^{\uparrow,\downarrow}_{ii}$  are real-valued by their definitions, while the correlated rates are complex in general.  We also have $\gamma_{ij}=\gamma_{ji}^*$. We refer to $ \gamma_{21}^{\downarrow,\uparrow}$ as dissipative couplings. Their real and imaginary parts stands for dissipative symmetric and DM interactions. Similar to the coherent DM interaction, here the dissipative DM interaction vanishes if the spacer layer is inversion symmetric by noting that (for example for $G\equiv \gamma_{21}^\downarrow$): 
\( \Im G=\frac{1}{2\hbar^2} \big[ \mathcal{G}^>_{21}(\Omega) +[ \mathcal{G}^>_{21}(\Omega) ]^*  \big] = \frac{1}{2\hbar^2} \big[  \mathcal{G}^>_{21}(\Omega) -\mathcal{G}^>_{12}(\Omega)  \big],  \)
where we have used the fact that $[ \mathcal{G}^>_{AB}(\omega) ]^*=- \mathcal{G}^>_{B^\dagger A^\dagger}(\omega) $.
 We also  remark that there is one important relation between the greater and lesser Green's functions
\(  \mathcal{G}^>_{ij}(\omega) = e^{\beta \hbar \omega} \mathcal{G}^<_{ij}(\omega),   \)
which implies that $\gamma^\downarrow_{ij}= e^{\beta \hbar \Omega} \gamma^\uparrow_{ji}$. This allows us to gauge away the phase of the correlated decay and absorption rates. Additionally, we note that, at zero temperature, the absorption rate becomes zero ($\mathcal{L}^\uparrow_{ij}\hat{\rho}(t)=0$), while the decay process persists. In this scenario, we obtain the master equation ({\color{red}2})  in the main text. We finally point out that the thermodynamic stability of the spacer imposes the constraint $\gamma^\uparrow_{jj} \geq |\gamma_{12}^\uparrow|$ and $\gamma^\downarrow_{jj} \geq |\gamma_{12}^\downarrow|$. This ensures that the dynamics of the system is  completely positive, while   Eq.~\eqref{s5} does not provide a general guarantee for complete positivity.

\subsection{(ii) Non-Hermtian Hamiltonian for the mean-field dynamics}
 In this section, we derive  the dynamics of the mean values of $\hat{m}_1, \hat{m}_2$, defined by $\langle \hat{m}_i \rangle \equiv \tr(\hat{\rho} \hat{m}_i)= \langle \hat{m}_i^x\rangle-i \langle \hat{m}_i^y\rangle $.  To this end, we only need to evaluate the following terms: 
\( \dv{\langle \hat{m}_i \rangle}{t} = -\frac{i}{\hbar}\langle [\hat{m}_i, \hat{H}]\rangle +\tr\big[ \hat{m}_i \sum_{k,j} \mathcal{L}_{kj}^\downarrow \hat{\rho} \big]  +\tr\big[ \hat{m}_i \sum_{k,j} \mathcal{L}_{kj}^\uparrow \hat{\rho} \big].  \)
We first consider the case of zero temperature, resulting in the vanishing of the last term in the equation above. In the following, we provide detailed derivation for the mode $\hat{m}_1$.
 The contribution from the coherent part is
\( -\frac{i}{\hbar} \langle [\hat{m}_1, \hat{H}] \rangle =-i\Omega \langle \hat{m}_1 \rangle -i \frac{\mathcal{J}}{\hbar} \langle \hat{m}_2 \rangle.  \)
In $\tr( \hat{m}_1 \sum \mathcal{L}^\downarrow_{kj}\hat{\rho} )$, we have four terms. For notational convenience, we introduce the local decay rate $\gamma\equiv \gamma^\downarrow_{jj}$ and the correlated decay rate $G\equiv \gamma^\downarrow_{21}$ (we adopt the same notations as in the main text for consistency and clarity).  The first term is 
\(  \gamma \tr\Big[ \hat{m}_1 \Big(  \hat{m}_1 \hat{\rho} \hat{m}_1^\dagger - \frac{1}{2}\{ \hat{m}_1^\dagger \hat{m}_1, \hat{\rho} \}    \Big)   \Big] = \gamma \Big( \langle \hat{m}_1^\dagger \hat{m}_1 \hat{m}_1  \rangle - \frac{1}{2} \langle \hat{m}_1 \hat{m}_1^\dagger \hat{m}_1 \rangle  - \frac{1}{2} \langle \hat{m}_1^\dagger \hat{m}_1 \hat{m}_1 \rangle \Big) = \frac{\gamma}{2} \langle [\hat{m}_1^\dagger, \hat{m}_1] \hat{m}_1\rangle= - \frac{\gamma}{2} \langle \hat{m}_1 \rangle.   \)
It describes the damping of $\langle \hat{m}_1\rangle$, where $\gamma$ is the local Gilbert damping. It is important to note that the effect of the quantum jump is not neglected in this treatment.
The second term is 
\(  \gamma \tr\Big[ \hat{m}_1 \Big(  \hat{m}_2 \hat{\rho} \hat{m}_2^\dagger - \frac{1}{2}\{ \hat{m}_2^\dagger \hat{m}_2, \hat{\rho} \}    \Big)   \Big] = \gamma \Big(  \langle \hat{m}_2^\dagger \hat{m}_2 \hat{m}_1  \rangle - \frac{1}{2}\langle \hat{m}_2^\dagger \hat{m}_2 \hat{m}_1  \rangle - \frac{1}{2}\langle \hat{m}_2^\dagger \hat{m}_2 \hat{m}_1  \rangle  \Big)=0,  \)
which implies that the local decay process of the second magnon mode does not impact the dynamics of $\langle \hat{m}_1\rangle$, as one may expect.  The third term is given by
\( G \tr\Big[ \hat{m}_1 \Big( \hat{m}_1 \hat{\rho} \hat{m}_2^\dagger - \frac{1}{2} \{ \hat{m}_2^\dagger \hat{m}_1, \hat{\rho} \}   \Big)   \Big] =G \Big( \langle \hat{m}_2^\dagger \hat{m}_1 \hat{m}_1 \rangle  - \frac{1}{2} \langle \hat{m}_2^\dagger \hat{m}_1 \hat{m}_1 \rangle - \frac{1}{2} \langle \hat{m}_2^\dagger \hat{m}_1 \hat{m}_1 \rangle   \Big)=0,  \)
and the fourth term is 
\( G^* \tr \Big[  \hat{m}_1 \Big( \hat{m}_2 \hat{\rho} \hat{m}_1^\dagger - \frac{1}{2} \{ \hat{m}_1^\dagger \hat{m}_2, \hat{\rho} \}   \Big)   \Big]  =  G^*  \Big(  \langle \hat{m}_1^\dagger \hat{m}_1 \hat{m}_2\rangle - \frac{1}{2} \langle \hat{m}_1 \hat{m}_1^\dagger \hat{m}_2 \rangle - \frac{1}{2} \langle \hat{m}_1^\dagger \hat{m}_1 \hat{m}_2  \Big)= \frac{G^*}{2} \langle [\hat{m}_1^\dagger, \hat{m}_1] \hat{m}_2 \rangle=- \frac{G^*}{2} \langle \hat{m}_2 \rangle.  \)
Therefore, the dynamics of the mean-field $\langle \hat{m}_1\rangle$ is governed by
\(  \dv{\langle \hat{m}_1\rangle}{t} = \Big( -i\Omega - \frac{\gamma}{2}  \Big)\langle \hat{m}_1\rangle +\Big( -i \frac{\mathcal{J}}{\hbar} - \frac{G^*}{2}  \Big) \langle \hat{m}_2\rangle , \)
and, similarly, the dynamics of $\langle \hat{m}_2\rangle$ is 
\(  \dv{\langle \hat{m}_2\rangle}{t} =  \Big( -i\Omega - \frac{\gamma}{2}  \Big)\langle \hat{m}_2\rangle    +\Big( -i \frac{\mathcal{J}^*}{\hbar} - \frac{G}{2}  \Big) \langle \hat{m}_1\rangle.   \)
We point out that they are the linearized Landau–Lifshitz–Gilbert equations. We can recast the two equations above into a more compact Schor\"{o}dinger-like equation with a non-Hermitian Hamiltonian: 
\( i\dv{\vec \psi}{t} = \mathcal{H}  \vec \psi, \)
with $\Vec{\psi}=(\langle \hat{m}_1\rangle, \langle \hat{m}_2\rangle  )^T$ and  
\(   \mathcal{H} = \mqty[ \Omega -i  \dfrac{\gamma}{2} &  \dfrac{\mathcal{J}}{\hbar}- i \dfrac{G^*}{2} \\  &    \\ \dfrac{\mathcal{J}^*}{\hbar} -i \dfrac{G}{2} & \Omega - \dfrac{i\gamma}{2}  ]. \label{eq:s23}  \)
This is the Hamiltonian ({\color{red}5}) in the main text. 

At finite temperature, we first have additional local terms in the master equation (for example for the first magnetic layer): 
\( \mathcal{L}_{11}^\uparrow =\gamma^\uparrow \Big[  \hat{m}_1^\dagger \hat\rho \hat{m}_1 - \frac{1}{2} \{ \hat{m}_1 \hat{m}_1^\dagger, \hat\rho \} \Big] .\)
We would like to highlight that this term can also account for the effect of local spin pumping, effectively reducing the local damping $\gamma$. This can be observed from its contribution to the equation of motion for $\langle \hat{m}_1\rangle$:
\( \tr(\hat{m}_1 \mathcal{L}_{11}^{\uparrow} [\hat\rho]) = \gamma^\uparrow \Big[  \langle \hat{m}_1 \hat{m}_1 \hat{m}_1^\dagger \rangle - \frac{1}{2}\langle \hat{m}_1 \hat{m}_1 \hat{m}_1^\dagger \rangle - \frac{1}{2} \langle \hat{m}_1 \hat{m}_1^\dagger \hat{m}_1\rangle  \Big] =\frac{\gamma^\uparrow }{2} \langle \hat{m}_1 [\hat{m}_1, \hat{m}_1^\dagger]\rangle=\frac{\gamma^\uparrow }{2} \langle \hat{m}_1\rangle.   \)
Then the equation for $\langle \hat{m}_1 \rangle$ is given by
\(  \dv{\langle \hat{m}_1\rangle}{t} = \Big( -i\Omega - \frac{\gamma-\gamma^\uparrow  }{2}  \Big)\langle \hat{m}_1\rangle +\Big( -i \frac{\mathcal{J}}{\hbar} - \frac{G^*}{2}  \Big) \langle \hat{m}_2\rangle , \label{eq:s26}  \)
where we see that the local damping $\gamma$ is reduced to $\gamma - \gamma^\uparrow$ as expected. At finite temperature, we also have the nonlocal terms (we focus on one specific term): 
\( \mathcal{L}^\uparrow_{21} \hat{\rho}=\gamma_{21}^\uparrow \big[   \hat{m}_1^\dagger \hat{\rho} \hat{m}_2 - \frac{1}{2} \{  \hat{m}_2 \hat{m}_1^\dagger, \hat{\rho}  \}   \big].   \)
This term would give a contribution: 
\(  \tr( \hat{m}_1  \mathcal{L}^\uparrow_{21} \hat{\rho} ) = \frac{ \gamma_{21}^\uparrow }{2} \langle \hat{m}_2\rangle , \)
which would shift 
\(   G^*\rightarrow G^*-  \gamma_{21}^\uparrow =(1-e^{-\beta\hbar\Omega}) G^* ,  \)
in Eq.~\eqref{eq:s26}. In the last step, we have used $\gamma^\uparrow_{ij}=e^{-\beta\hbar \Omega}\gamma_{ji}^\downarrow$ and $(\gamma^{\uparrow,\downarrow}_{ij})^*=\gamma^{\uparrow,\downarrow}_{ji}$.
Here, we would like to emphasize that the introduction of finite temperature does not alter the structure of the Hamiltonian~\eqref{eq:s23} but it does lead to a reduction in the values of the local damping $\gamma$ and the dissipative coupling $G$. On the other hand, since the dissipative coupling is a crucial component for our proposal of a dissipative spin-wave diode, it is necessary to ensure that the temperature is lower than the resonance frequency, i.e., $k_BT\leq \hbar \Omega$, in order to maintain a finite effective dissipative coupling.

\subsection{(iii) Scattering matrix for the spin-wave diode}
To demonstrate the realization of a spin-wave diode in our setup, we incorporate magnon waveguides (such as YIG films) at each end, as depicted in Fig. 1(a) in the main text. By assuming a coupling rate of $\kappa$ between the macrospin and the waveguide, we need to replace the local damping $\gamma$ with $\Gamma=\gamma +\kappa$.
 In this case, the Hamiltonian reads: 
\(   \mathcal{H} = \mqty[ \Omega -i  \Gamma/2 &  \mathcal{J}/\hbar - i {G^*}/{2} \\  &    \\ {\mathcal{J}^*}/{\hbar} -i {G}/{2} & \Omega - i {\Gamma}/{2}  ].  \)
 The corresponding Green's function is given by: 
\( \hat{G}(\omega)= \frac{1}{\det[\omega-\mathcal{H}]} \mqty[  \omega-\Omega + i\frac{\Gamma}{•2}  &   \frac{\mathcal{J}}{\hbar} - i \frac{G^*}{2}  \\ \frac{\mathcal{J}^*}{\hbar} -i \frac{G}{2}  & \omega-\Omega +i \frac{\Gamma}{2}    ]   ,  \label{eq:s32}  \)
with $\det[\omega-\mathcal{H}] = (\omega-\Omega + i\Gamma/2)^2  -(\mathcal{J}/\hbar - i G^*/2 ) ( \mathcal{J}^*/\hbar -iG/2  ).$  The scattering matrix $\hat{S}(\omega)$ is related to the Green's function  according to input-output theory:
\(   \hat{S}(\omega)=1-i\kappa \hat{G}(\omega).  \)
With the obtained expression for the Green's function, we can readily derive the scattering matrix. For simplicity,  we assume the dissipative coupling $G$ to be positive, with the phase absorbed into the definition of $\mathcal{J}$, as done in the main text, and we assume $\mathcal{J}=e^{i\Phi}|\mathcal{J}|$.  In this analysis, we consider two cases. 

\textbf{Case I.}|We consider the scenario where the spin wave is at resonance, i.e., $\omega=\Omega$. In this case, the scattering matrix can be expressed as follows:
\( \hat{S}(\Omega)=I_{2\times 2} + \frac{i\kappa}{ \Gamma^2/4 +(\mathcal{J}/\hbar -i G/2) (\mathcal{J}^*/\hbar-iG/2) } \mqty[ i\Gamma/2 & \mathcal{J}/\hbar-iG/2 \\ \mathcal{J}^*/\hbar -iG/2 & i\Gamma/2 ].   \)
We first set  $\Phi=\pi/2$ thus $\mathcal{J}=i|\mathcal{J}|$ (the coherent coupling is purely DM interaction). We also assume the local damping equals the dissipative coupling $\gamma=G$ for simplicity. Then the two transmission coefficients are given by
\(  | \hat{S}_{12}(\Omega) |^2= \Bigg[   \frac{(\kappa/\gamma) (|\mathcal{J}|/G\hbar -1/2  ) }{ (\kappa/\gamma+1)^2/4 +\big[ ( |\mathcal{J}|/G\hbar  )^2  -1/4 \big]   }    \Bigg]^2, \;\;\;  | \hat{S}_{21}(\Omega) |^2= \Bigg[   \frac{(\kappa/\gamma) (|\mathcal{J}|/G\hbar +1/2  ) }{ (\kappa/\gamma+1)^2/4 +\big[ ( |\mathcal{J}|/G\hbar  )^2  -1/4 \big]   }    \Bigg]^2,        \)
from which we plot Fig.~2(a) and (b) in the main text. To examine the effect of the phase $\Phi$, we set $|\mathcal{J}|/\hbar=G/2$ and assume $\kappa=\gamma=G$. In this case, we have
\(  | \hat{S}_{12}(\Omega) |^2= \Bigg|  \frac{2 (e^{i\Phi} -i )  }{4 +(e^{i\Phi} -i )(e^{-i\Phi} -i ) }  \Bigg|^2,\;\;\; | \hat{S}_{21}(\Omega) |^2= \Bigg|  \frac{2 (e^{-i\Phi} -i )  }{4 +(e^{i\Phi} -i )(e^{-i\Phi} -i ) }  \Bigg|^2,   \)
from which we plot the Fig.~2(c) in the main text. 

\textbf{Case II.}| Here, we consider the case where the optimal condition is satisfied: $\mathcal{J}/\hbar=iG/2$. In this case, $\hat{S}_{12}(\omega)=0$ for all frequencies. We assume $\gamma=\kappa$. In this case, we have: 
\( | \hat{S}_{21}(\omega)|^2 =  \frac{(G/\gamma)^2}{ [ 1+ ( \omega-\Omega )^2/\gamma^2 ]^2  }, \;\;\; |\hat{S}_{ii}(\omega)|^2= \frac{   ( \omega-\Omega )^2/\gamma^2   }{1+ ( \omega-\Omega )^2/\gamma^2} . \)
From these two expressions, we plot Fig.~2(d) in the main text. When we pump the system, we replace $\gamma$ with $\tilde{\gamma}=\gamma-\gamma^\uparrow$. Then we obtain the Fig.~3 in the main text. To determine the regime of amplification, we set $ | \hat{S}_{21}(\omega)|^2 >1$, yielding: 
\(  |\omega-\Omega|<\sqrt{ (G-\tilde{\gamma}) \tilde{\gamma}  },  \)
which is the dashed line in Fig.~3 in the main text. We note that $G$ is always larger than $\tilde{\gamma}$ in the amplification regime.

\subsection{ (iv) Topological index and unidirectional amplification}
When we generalize our setup to a ring, the Hamiltonian in momentum space reads: 
\(  \mathcal{H}=\sum_k h(k) \hat{m}_k^\dagger \hat{m}_k, \;\;\;\text{with}\;\;\; h(k)=\hbar(\Omega-i\gamma)+(\mathcal{J}-i\hbar G/2)e^{ik} +(\mathcal{J}^*-i\hbar G/2)e^{-ik},     \)
as we discussed in the main text. The central quantity is the Green's function: $\hat{G}(\omega)=(\omega -\mathcal{H}/\hbar)^{-1}$, which can be evaluated by
                  \( \bra{j}\hat{G}(\omega)\ket{l} = \int^{2\pi}_0 \frac{dk}{2\pi}   \frac{e^{ik(j-l)}}{\omega-h(k)/\hbar}  =\oint_{|z|=1} \frac{dz}{2\pi i} \; \frac{z^{j-l-1}}{\omega - h(z)/\hbar}.   \)
                   Here, we have converted the line integral to a complex integral (by setting $z=e^{ik}$), which can be easily evaluated using Cauchy's residue theorem.  The function $h(z)$ is given by $h(z)=\hbar(\Omega-i\gamma)+(\mathcal{J}-i\hbar G/2)z+(\mathcal{J}^*-i\hbar G/2)z^{-1}$. The case of interest is when the condition $\mathcal{J}=i\hbar G/2$ is satisfied. In this case, we have a perfect diode and the function $h(z)$ is reduced to $h(z)=\hbar(\Omega-i\gamma)-i\hbar G z^{-1}$. Just for sanity check, we can evaluate the Green' function for two adjacent sites: 
           \( \al{ & \bra{j}\hat{G}(\omega)\ket{j} =\oint_{|z|=1} \frac{dz}{2\pi i} \frac{1}{z[\omega - h(z)/\hbar]} = \frac{1}{\omega-\Omega +i\gamma} \oint \frac{dz}{2\pi i} \frac{1}{z+ 2iG/(\omega-\Omega+i\gamma)} = \frac{1}{\omega-\Omega +i\gamma}, \\
& \bra{j}\hat{G}(\omega)\ket{j+1}  =\oint_{|z|=1} \frac{dz}{2\pi i} \frac{1}{z^2[\omega - h(z)/\hbar]} =  \frac{1}{\omega-\Omega +i\gamma} \oint \frac{dz}{2\pi i} \frac{1}{z\big[z+ iG/(\omega-\Omega+i\gamma)\big]} = 0,\\
& \bra{j+1}\hat{G}(\omega)\ket{j}=\oint_{|z|=1} \frac{dz}{2\pi i} \frac{1}{\omega - h(z)/\hbar} =  \frac{1}{\omega-\Omega +i\gamma} \oint \frac{dz}{2\pi i} \frac{z}{z+ iG/(\omega-\Omega+i\gamma)} = \frac{-iG}{(\omega - \Omega+i\gamma)^2}. }\)    
This reproduces the Green's function in Eq.~\eqref{eq:s32}, taking into account that the local damping is $2\gamma$ as each magnetic layer is adjacent to two spacer layers. In general, we have the following result (for $l>0$): 
\(   \bra{j}\hat{G} (\omega)\ket{j+l}=0,\;\;\; \text{and} \;\;\;    \bra{j+l}\hat{G}(\omega)\ket{j} =     \oint_{|z|=1} \frac{dz}{2\pi i}\; \frac{z^{l-1}}{\omega-h(z)/\hbar}= \bigg( \! \frac{-iG}{\omega-\Omega +i\gamma }  \!\bigg)^l   \frac{1}{\omega-\Omega +i\gamma},    \)
when $G<|\omega-\Omega+i\gamma|$.
In this case, the launched spin wave propagating in the clockwise direction is blocked, while the wave propagating in the counterclockwise direction decays exponentially. When $G>|\omega-\Omega+i\gamma|$, the value of $ \bra{j+l}\hat{G}(\omega)\ket{j} $ abruptly transitions to zero (the single pole of the integrand is outside the unit circle). This phenomenon is a consequence of the system entering the amplification phase (unstable in the ring structure). Given the system's ring structure, the spin wave continuously accumulates energy, leading to instability. To distinguish these two phases, we can also assign a topological index to them. To this end, we introduce a planar vector field:
 \( \vb n(k) = \Big( \Im{\omega- h(k)/\hbar}, \Re{\omega -h(k)/\hbar}  \Big) = \big( \gamma+G\cos k, \omega-\Omega+G\sin k  \big). \)
  We can associate a winding number with this vector field: 
\( \mathcal{N}=\frac{1}{2\pi} \int_0^{2\pi} dk\; \hat{z}\cdot \vb n \times \partial_k \vb n.  \)
\begin{figure}
	\centering\includegraphics[width=\linewidth]{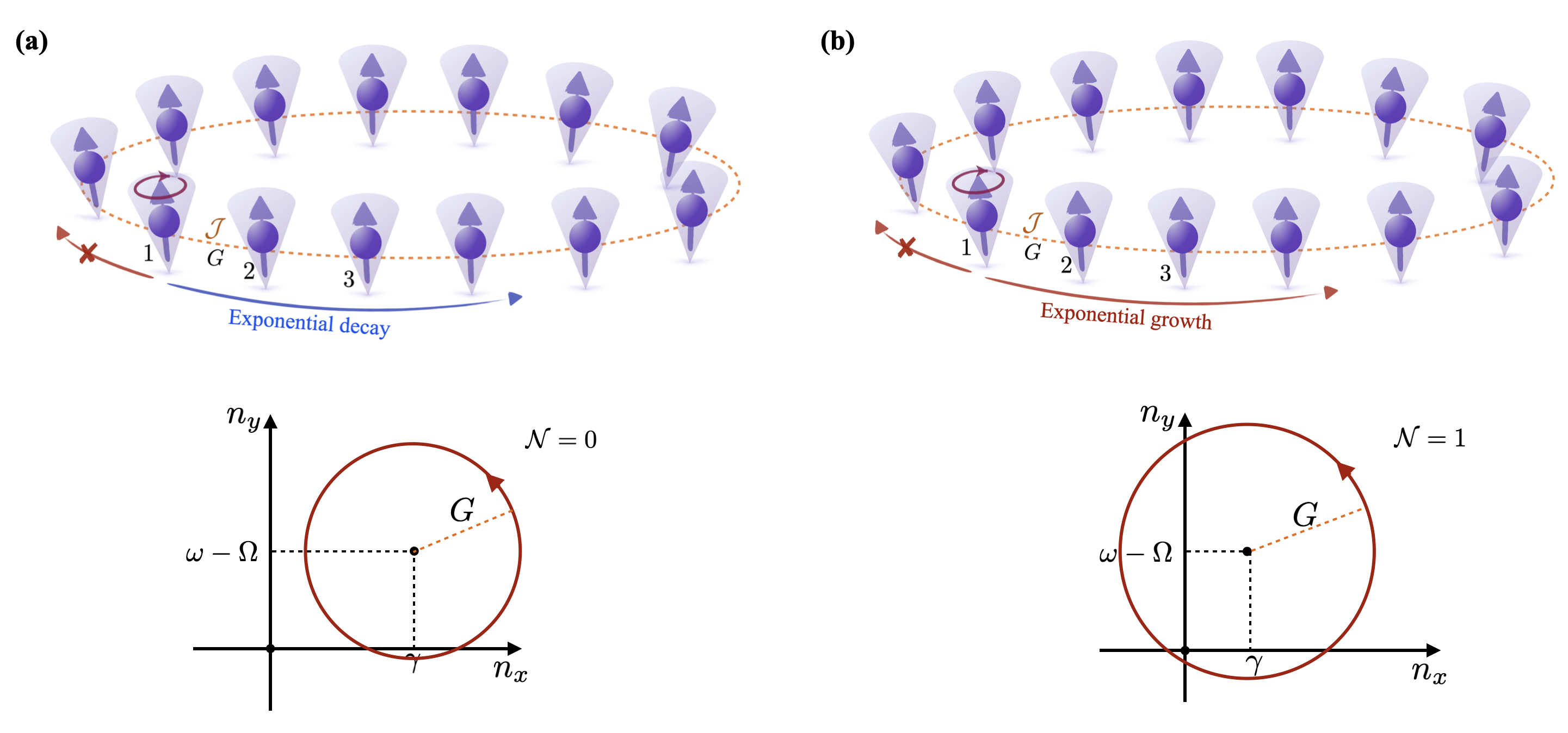}
 \caption{(a) The winding number $\mathcal{N}=0$ corresponds to a small dissipative coupling $G^2<(\omega-\Omega)^2+\gamma^2$. In this case, the clockwise-propagating spin wave is blocked, while the counterclockwise-propagating wave decays exponentially.
(b) The winding number $\mathcal{N}=1$ corresponds to a large dissipative coupling $G^2>(\omega-\Omega)^2+\gamma^2$. In this scenario, the counterclockwise-propagating spin wave grows exponentially, leading to system instability. }
  \label{sm1}
\end{figure}
As shown in Fig.~\ref{sm1}(a), when $G^2<(\omega-\Omega)^2+\gamma^2$, the winding number $\mathcal{N}$ is zero ($\mathcal{N}=0$) and the launched spin wave exhibits exponential decay. Conversely, when $G^2>(\omega-\Omega)^2+\gamma^2$, the winding number $\mathcal{N}$ equals one ($\mathcal{N}=1$)  as indicated in Fig.~\ref{sm1}(b), and the spin wave undergoes exponential growth, thereby inducing system instability. We note that we have assumed $\mathcal{J}=i\hbar G/2$. However, we can also choose $\mathcal{J}^*=i\hbar G/2$; then the counterclockwise-propagating launched spin wave would be blocked. In this scenario, the winding number in the unidirectional amplification phase would be $\mathcal{N}=-1$.

\end{document}